\documentstyle[prb,aps]{revtex}

\begin{document}
\draft
\title{Kondo effect in a quantum critical ferromagnet}
\author{Yupeng Wang} 
\address{
Institut f\"{u}r Physik, Universit\"{a}t Augsburg, D-86135, Augsburg, Germany\\
and Laboratory of Ultra-Low Temperature Physics, Chinese Academy of Sciences, 
\\ Beijing 100080,  People's Republic of China}
\author{Jianhui Dai}
\address{
Abdus Salam International Center for Theoretical Physics, Trieste 34100, Italy \\ and
Zhejiang Institute of Modern Physics, Zhejiang University,\\ Hangzhou 310027, People's Republic of China}
\maketitle
\begin{abstract}
We study the Heisenberg ferromagnetic spin chain coupled with a boundary impurity. Via Bethe ansatz
solution, it is found that (i) for $J>0$, the impurity
spin behaves as a diamagnetic center and is completely screened by $2S$ bulk spins in the ground state, 
no matter how large the impurity spin 
is; (ii) the specific heat of the local composite (impurity plus $2S$ bulk spins which form bound state
with it) shows a simple power law $C_{loc}\sim T^{\frac32}$;  (iii)for $J<0$, the impurity is locked into the critical behavior of the
bulk. Possible phenomena in higher dimensions are discussed.  
\end{abstract}
\pacs{75.10.Jm, 75.20.Hr, 72.15.Qm, 05.70.Jk}
Kondo problem or the magnetic impurity problem in an electron host plays a very important role in 
modern condensed matter physics. It represents a generic non-perturbationable example of the strongly
correlated many-body systems. Recently, with the development of research on some low-dimensional
systems\cite{1} and the observation of unusual non-Fermi-liquid behavior in some heavy fermion
compounds\cite{2}, the interest in this problem has been largely renewed. The multi-channel Kondo
problem\cite{3} provided the first example of impurity systems which show non-Fermi-liquid behavior
at low temperatures\cite{4}. In a Luttinger liquid, the impurity behaves rather different\cite{5,6}
from that in a Fermi liquid and  may interpolate between a local Fermi liquid and some non-Fermi
liquid.\cite{7}  Some new quantum critical phenomena have also been predicted in some integrable 
models\cite{8,9}. Generally speaking, these new findings indicate that the quantum 
impurity models renormalize to critical points
 corresponding to conformally invariant boundary conditions\cite{10}. 
 Another important progress is the study on the Kondo problem in Fermi systems with
pseudo gap\cite{11}, i.e., the density of states $\rho(\epsilon)$ is power-law-dependent on the energy,
$\rho(\epsilon)\sim \epsilon^r$. With renormalization group (RG) analysis, Withoff and Fradkin\cite{11} 
showed that there is a critical value $J_c$ for the Kondo coupling constant $J$. For $J>J_c$, Kondo
 effect occurs at low temperatures, while for $J<J_c$, the impurity decouples from the host. 
We note that all the quantum critical behaviors mentioned above only occur for $T\to 0$ and therefore
fall into the general category of quantum phase transitions\cite{12}.
\par
In an earlier publication, Larkin and Mel'nikov studied the Kondo effect in an 
almost ferromagnetic metal\cite{13}. With the traditional perturbation theory they
showed that the impurity susceptibility is almost Curie type with logarithmic corrections
at intermediately low temperatures. However, the critical behavior of a Kondo impurity
in a quantum critical ferromagnet has never been touched. The main difficulty in approaching this
problem is that almost all perturbation techniques fail in the critical regime and exact results
are expected. As discussed in some recent
works\cite{14}, the critical behavior of the impurity strongly depends  on the host properties
and seems to be non-universal. Typical quantum critical
ferromagnet is the Heisenberg system in reduced dimensions ($d\leq 2$). These
systems have long-range-ordered ground states but are disorder at any finite
temperatures due to the strong quantum fluctuations. In this paper, 
we study the critical behavior of an impurity spin coupled with a
Heisenberg ferromagnetic chain. 
The model Hamiltonian we shall consider
reads
\begin{eqnarray}
H=-\frac12\sum_{j=1}^{N-1}{\vec \sigma}_j\cdot{\vec \sigma}_{j+1}+J{\vec \sigma}_1\cdot{\vec S},
\end{eqnarray}
where ${\vec \sigma}_j$ is the Pauli matrices on site $j$; $N$ is the length of the 
chain; ${\vec S}$ is the impurity spin sited at one end of the chain; $J$ is a real 
constant which describes the Kondo coupling between the impurity and the host.
The problem is interesting because (i)the model is not conformally
invariant due to the nonlinear dispersion relation of the low-lying excitations, $\epsilon(k)\sim k^2$,
and $\rho(\epsilon)\sim\epsilon^{-\frac12}$, and represents a typical quantum critical system beyond
the universality of the conventional Luttinger liquid;
(ii)the Hamiltonian is very simple
(without any superfluous term) and  allows exact solution via algebraic Bethe 
ansatz\cite{15}. In fact, most known methods\cite{8,9} developed for the impurity problem in
a Luttinger liquid can not be used for the present system due to the strong quantum fluctuations.
\par
Let us first summerize the solution of (1). Define the Lax operator
$L_{j\tau}(\lambda)\equiv \lambda+i/2(1+{\vec \sigma}_j\cdot{\vec \tau})$, where
${\vec \tau}$ is the Pauli matrices acting on the auxiliary space and $\lambda$ is
the so-called spectral parameter. For the impurity, we define $L_{0\tau}\equiv
\lambda+i(1/2+{\vec S}\cdot{\vec \tau})$. Obviously, $L_{j\tau}$ and $L_{0\tau}$
satisfy the Yang-Baxter equation (YBE)\cite{16}. It can be easily shown that the
doubled-monodromy matrix
\begin{eqnarray}
T_\tau(\lambda)\equiv L_{N\tau}(\lambda)\cdots L_{1\tau}(\lambda)
L_{0\tau}(\lambda-ic)L_{0\tau}(\lambda+ic)L_{1\tau}(\lambda)\cdots L_{N\tau}(\lambda),
\end{eqnarray}
satisfies the reflection equation
\begin{eqnarray}
L_{\tau\tau'}(\lambda-\mu)T_\tau(\lambda)L_{\tau\tau'}(\lambda+\mu)T_{\tau'}(\mu)
=T_{\tau'}(\mu)L_{\tau\tau'}(\lambda+\mu)T_\tau(\lambda)L_{\tau\tau'}(\lambda-\mu).
\end{eqnarray}
From the above equation we can show that the transfer matrices $\theta(\lambda)\equiv
Tr_\tau T_\tau(\lambda)$ with different spectral parameters are commutative,
$[\theta(\lambda),\theta(\mu)]=0$. Therefore, $\theta(\lambda)$ serves as a generator
of a variety of conserved quantities. The Hamiltonian Eq.(1) is given by
\begin{eqnarray}
H=\frac i2J(-1)^N\frac{\partial}{\partial \lambda}\theta(\lambda)|_{\lambda=0}+
\frac12(N+1-J),
\end{eqnarray}
with $J=1/{[c^2-(S+1/2)^2]}$. Following the standard method\cite{15,8} we obtain the Bethe ansatz equation (BAE) 
\begin{eqnarray}
\left(\frac{\lambda_j-\frac i2}{\lambda_j+\frac i2}\right)^{2N}\frac{\lambda_j
-i(S+c)}{\lambda_j+i(S+c)}\frac{\lambda_j-i(S-c)}{\lambda_j+i(S-c)}
=\prod_{l\neq j}^M\frac{\lambda_j-\lambda_l-i}{\lambda_j-\lambda_l+i}
\frac{\lambda_j+\lambda_l-i}{\lambda_j+\lambda_l+i},
\end{eqnarray}
with the eigenvalue of Eq.(1) as
\begin{eqnarray}
E(\{\lambda_j\})=\sum_{j=1}^M\frac{1}{\lambda_j^2+\frac14}-\frac12(N-1)+JS,
\end{eqnarray}
where $\lambda_j$ represent the rapidities of the magnons and M the number of the magnons.
\par
{\sl Ground state}. In the thermodynamic limit, the bulk solutions of $\lambda_j$ are described
by the so-called n-strings\cite{17}. However, due to the presence of the impurity,
some boundary bound states may exist for $c>S$, which are usually called  the
$n-k$-strings\cite{18}:
\begin{eqnarray}
\lambda_b^m=i(c-S)+im, {~~~~}m=k, k+1, \cdots n.
\end{eqnarray}
In the ground state, only some $n-0$-strings may survive. We call them 
boundary $n$-strings. In our case, $n\geq 0$ has also an upper bound $n\leq 2S-1$
since $\lambda_j=\pm i(c+S)$ are forbidden as we can see from Eq.(5). No bulk
strings can exist at zero temperature since they carry positive energy. Boundary
bound state can exist only for $c>S+1/2$ (antiferromagnetic Kondo coupling)
because in this case, the boundary $n$-strings carry negative energy. For zero
external magnetic field, the most stable boundary string has the length of $2S$ 
with the energy $\epsilon_{2S}=2S/[S^2-(c-1/2)^2]$. Therefore, the impurity contributes
a magnetization of $-S$. Such a phenomenon can be understood in a simple picture.
Due to the antiferromagnetic coupling between the impurity and the bulk, $2S$ bulk
spins are swallowed by the impurity at zero temperature to form a $2S+1$-body
singlet. This singlet does not contribute to the magnetization of the ground state.
In this sense, the impurity is completely screened, no matter how large the impurity
moment is. Such a situation is very different from that of the conventional Kondo 
problem, where the impurity moment can only be partially screened by the host when
$S>S'$ ($S'$ the spin of the host particles)\cite{19}. This difference is certainly due
to the different properties of the hosts. In the antiferromagnetic spin chain or a
normal metal, the spin correlation of the bulk is antiferromagnetic type which repels more than one
bulk spin or electron to screen the impurity. However, in a ferromagnetic spin chain, the
bulk correlation is ferromagnetic which allows and in fact enhances some bulk spins
to form a larger moment to screen the impurity. The local singlet is nothing but
a bound state of $2S$ magnons. The boundary string may be 
broken by the external field. In fact, there are $2S$ critical fields
\begin{eqnarray}
H_c^n=\frac1n\left[\frac {2S}{(c-\frac12)^2-S^2}-\frac{2S- n}{(c-\frac{n+1}2)^2
-(S-\frac n2)^2}\right], {~~~~}n=1,2,\cdots,2S.
\end{eqnarray}
When $H_c^{n}<H<H_c^{n+1}$, only a boundary $(2S-n)$-string survives in the ground 
state and when $H>H_c^{2S}$, any boundary string becomes unstable. Notice that at
$H=H_c^n$, the ground-state-magnetization has a jump $\delta M=1$, which corresponds
to some type of quantum phase transition. The finite value of $H_c^1$ indicates that the zero 
temperature susceptibility of the local singlet is exactly zero.
\par
{\sl Thermal BAE}. Since we are interested mostly in the critical behavior, we consider $T,H<< H_c^1$
and $J>0$ case in the following text. In this case, any excitations breaking the boundary string
can be plausibly omitted due to the energy gap associated with them.
 With the standard thermal Bethe ansatz\cite{17}, we derive the
thermal BAE as
\begin{eqnarray}
\ln(1+\eta_n)=\frac{2\pi a_n(\lambda)+nH}T+\sum_{m=1}^\infty {\bf A}_{mn}\ln[1+\eta_m^{-1}(\lambda)],
\end{eqnarray}
or equivalently,
\begin{eqnarray}
\ln\eta_1(\lambda)=\frac\pi Tg(\lambda)+{\bf G}\ln[1+\eta_2(\lambda)],\nonumber\\
\ln\eta_n(\lambda)={\bf G}\{\ln[1+\eta_{n+1}(\lambda)]+\ln[1+\eta_{n-1}(\lambda)]\}, {~~~}
n>1,\\
\lim_{n\to\infty}\frac{\ln\eta_n}n=\frac HT\equiv 2x_0,\nonumber
\end{eqnarray}
where $a_n(\lambda)=n/2\pi[\lambda^2+(n/2)^2]$, ${\bf A}_{mn}=[m+n]+2[m+n-2]+\cdots+2[|m-n|+2]
+[|m-n|]$; $g(\lambda)=1/2\cosh(\pi\lambda)$; $\eta_n(\lambda)$ are some functions which determine
the free energy of the system; and $[n]$ and ${\bf G}$ are integral operators 
with the kernels $a_n(\lambda)$
and $g(\lambda)$, respectively.  The free energy is given by
\begin{eqnarray}
F=F_{bulk}+F_{imp},\nonumber\\
F_{bulk}=F_0-(N+\frac12)T\int g(\lambda)\{\ln[1+\eta_1(\lambda)]-\frac{2\pi a_1(\lambda)+H}T\}d\lambda,\\
F_{imp}=\frac12T\sum_{n=1}^\infty\int\phi_n'(\lambda)\ln[1+\eta_n^{-1}(\lambda)]d\lambda,\nonumber
\end{eqnarray}
where $a_{n,m}(\lambda)=\sum_{l=1}^{min(m,n)}a_{n+m+1-2l}(\lambda)$; $\phi_n'(\lambda)=
a_{n,2S}(\lambda-ic+i)+a_{n,2S}(\lambda+ic-i)$; $F_0$ is the 
ground state energy; $F_{bulk}$ and $F_{imp}$ are the free energies of the bulk (including the bare
boundary) and the impurity,
respectively.
Notice that Eq.(9) and Eq.(10) are more
difficult to handle than those of the antiferromagnetic chain\cite{17}, since here all
$\eta_n$ diverge as for $T\to 0$. These equations were solved numerically\cite{20}
in studying  the critical behavior of the ferromagnetic Heisenberg chain. In 
addition, Schlottmann gave an analytical result based on a simple correlation-length
approximation\cite{21} and the result coincides with the numerical ones very
well. As we can see
from Eq.(9) and Eq.(10), when $T\to 0$, $\eta_n\to\infty$. To arrive at  the asymptotic solutions of 
$\eta_n(\lambda)$, we make the ansatz $\eta_n(\lambda)=\exp[2\pi a_n(\lambda)/T]\phi_n$.
Substituting this ansatz into Eq.(10) we readily obtain $\phi_n\sim1$ for finite $n$ and $\lambda$.
Therefore,
\begin{eqnarray}
\eta_n\approx\exp[\frac{2\pi a_n(\lambda)}T], {~~~~~~~~}T\to 0.
\end{eqnarray}
On the other hand, when $\lambda\to\infty$ or $n\to\infty$, the driving term in Eq.(10) tends to zero. 
This gives another asymptotic solution of $\eta_n$ for very large $\lambda$ or $n$\cite{17}
\begin{eqnarray}
\eta_n=\frac{\sinh^2[(n+1)x_0]}{\sinh^2x_0}-1+O(\frac1Te^{-\pi|\lambda|}),
\end{eqnarray}
For intermediate $\lambda$ and $n$ we have a crossover regime. We call Eq.(12) as the strong-coupling 
solution, while Eq.(13) as the weak-coupling solution.
 By equating them we obtain
two types of crossover scales, $\lambda_c(n)$ for small $n$ and $n_c(T)$,
\begin{eqnarray}
\lambda_c(n)\approx \left[\frac n{4T\ln(1+n)}\right]^{\frac12},{~~~~~~}n_c(T)\approx
\frac1{4T\ln(1+n_c)}\approx-\frac1{4T\ln T},
\end{eqnarray}
which characterize the crossover of the strong-coupling regime and the weak-coupling regime. 
Notice that the strong-coupling solution gives the correct ground state energy and the low-temperature
thermodynamics is mainly dominated by the weak-coupling solution\cite{22}. With
such an approximation, the recursion for $\eta_n$ can be performed by substituting the asymptotic
solutions into the right hand side of Eq.(9) and therefore the leading order correction upon the
asymptotic solutions can be obtained. In the following recursion process, we adopt the strong-coupling
 solution in the region of $\lambda<\lambda_c$ and $n<n_c$, while the weak-coupling solution is adopted
in other cases. This corresponds to an abrupt crossover, which does not affect the temperature 
dependence of the thermodynamic quantities in leading orders 
but their amplitudes.  For convenience, we define $\zeta_n(\lambda)\equiv\ln[1+\eta_n(\lambda)]
-[2\pi a_n(\lambda)+nH]/T$, which are responsible for the temperature-dependent part of the free energy.
\par
{\sl Low-temperature susceptibility of the impurity}. For convenience, we consider $2c=integer$ case.
Taking the boundary string into account, the free energy of the impurity can be rewritten as
\begin{eqnarray}
F_{imp}=\frac12T\int g(\lambda)[\zeta_{2c+2S-2}(\lambda)-sgn(2c-2S-2)\zeta_{|2c-2S-2|}(\lambda)]d\lambda.
\end{eqnarray}
Substituting the asymptotic solutions Eq.(12) and Eq.(13) into Eq.(9) and omitting the exponentially small terms,
we obtain
\begin{eqnarray}
\zeta_n(\lambda)\approx\sum_{m=1}^{n_c}\{\ln[1+\frac1{m(m+2)}]-\frac23x_0^2\}[\int_{\lambda_c(m)}^\infty 
+\int_{-\infty}^{-\lambda_c(m)}]A_{mn}(\lambda-\lambda')d\lambda'\nonumber\\
+2n_c\ln\frac{\sinh(1+n_c)x_0}{\sinh n_cx_0},
\end{eqnarray}
where $A_{mn}$ is the kernel of ${\bf A}_{mn}$. For small $n<<n_c$, up to the leading order, 
we find that the $x_0^2$ term of $\zeta_n(\lambda)$ is exactly $n$-times of that of 
$\zeta_1(\lambda)$. 
From Eq.(15) we easily derive
\begin{eqnarray}
\chi_{imp}=-2S\chi_{bulk}+subleading{~}order{~}terms,
\end{eqnarray}
where $\chi_{bulk}\sim T^{-2}\ln^{-1}(1/T)$ is the per-site susceptibility of the bulk\cite{20,21}.
Very interestingly, the impurity contributes a negative susceptibility, which indicates a novel
Kondo diamagnetic effect. That means the Kondo coupling dominates always over the ``molecular field"
generated by the bulk ferromagnetic fluctuations. Notice that Eq.(17) is only the contribution
of the bare impurity. If we take the screening cloud ($2S$ bulk spins which form the 
bound state with the impurity) into account, we find that the total susceptibility of the local
singlet is exactly canceled in the leading order. That means the polarization effect of the local
bound state only occurs in some subleading order, which indicates a strong coupling fixed point
$J^*=\infty$. In fact, the local singlet is much more insensitive to a small external magnetic
field as we discussed for the ground state. When $T\to 0$, its susceptibility must tend to
zero due to the bound energy as shown in Eq.(8). We note the present method is not reliable
to derive the total susceptibility of the local singlet but the above picture must be true.
The same conclusion can be achieved for arbitrary $J>0$. 
\par
{\sl Specific heat of the local composite}. In the framework of the local Fermi-liquid theory\cite{23}, the Kondo effect is 
nothing but the scattering effect of the rest bulk particles ($N-2S$) off the local-spin-singlet
composite or equivalently, the polarization effect of the local composite due
to the scattering. Taking the boundary string into account, the BAE of the bulk
modes can be rewritten as
\begin{eqnarray}
\left(\frac{\lambda_j-\frac i2}{\lambda_j+\frac i2}\right)^{2(N-2S)}=e^{i\phi(\lambda_j)}
\prod_{l\neq j}^{M-2S}\frac{\lambda_j-\lambda_l-i}{\lambda_j-\lambda_l+i}
\frac{\lambda_j+\lambda_l-i}{\lambda_j+\lambda_l+i},\\
e^{i\phi(\lambda)}=\frac{\lambda-i(c+S-1)}{\lambda+i(c+S-1)}\frac{\lambda+i(c-S-1)}
{\lambda-i(c-S-1)}\left(\frac{\lambda+\frac i2}{\lambda-\frac i2}\right)^{4S},
\end{eqnarray}
where $\phi(\lambda)$ represents the phase shift of a spin wave scattering off 
the local composite (boundary bound state). 
When $S=1/2$, $c\to 1+0^+$ or $J\to +\infty$, $\phi(\lambda)=0$. That means one of the bulk
spin is completely frozen by the impurity and the system is reduced to an $N-1$-site 
ferromagnetic chain. When $S=1/2$, $1<c<3/2$, 
only $\zeta_1(\lambda)$ is relevant and the free energy of the
local composite reads
\begin{eqnarray}
F_{loc}=-T\int g(\lambda)[\zeta_1(\lambda)-\frac12\zeta_1(\lambda-ic+i)-\frac12
\zeta_1(\lambda+ic-i)]d\lambda.
\end{eqnarray}
When $x_0=0$, we have
\begin{eqnarray}
\zeta_1(\lambda)-\frac12\zeta_1(\lambda-ic+i)-\frac12
\zeta_1(\lambda+ic-i)\nonumber\\
=16(c-1)^2T^{\frac32}\frac1\pi\sum_{m=1}^{n_c}\ln[1+\frac1{m(m+2)}]
m^{-\frac12}\ln^{\frac32}(1+m)+\cdots.
\end{eqnarray}
The sum in the above equation is convergent for large $n_c$.
Therefore we can extend it to infinity, which gives
the low-temperature specific heat of the local composite as
\begin{eqnarray}
C_{loc}\sim T^{\frac32}.
\end{eqnarray}
Similar conclusion can be arrived for arbitrary $S$ and $J>0$.
As long as the Kondo coupling is antiferromagnetic ($c\geq S+1/2$), the low-temperature
specific heat 
of the local composite is described by Eq.(22). There is a slightly difference
between $S=1/2$ case and $S>1/2$ case. For the former when $J\to\infty$, the local
singlet is completely frozen and $C_{loc}\to 0$, while for the later even when
$J\to\infty$, $C_{loc}$ takes a finite value. This can be
understood in a simple picture. For $S>1$, more than one bulk spin will be trapped by the impurity. Even
for $J\to\infty$, only one bulk spin (on the nearest neighbor site) can be completely frozen and the
rest is still polarizable via the bulk fluctuation. We note the specific heat of the local singlet is
much weaker than that of the Kondo impurity in a conventional metal. This still reveals the insensitivity
of the local bound state to the thermal activation. Though the anomalous power law Eq.(22) looks very 
like that obtained in the Luttinger Kondo systems\cite{8,9}, they are induced by different mechanisms.
In the present case, this anomaly is mainly due to the strong quantum fluctuation while in the Luttinger
liquid, the anomaly is in fact induced by the tunneling effect of the conduction electrons through the
impurity\cite{6,24,25}.
\par
For the ferromagnetic coupling case ($J<0$), no boundary bound state exists. Even in the ground state,
the impurity spin is completely polarized by the bulk spins. At finite temperature, the critical behavior
is locked into that of the bulk ($C_{imp}\sim T^{\frac12}$, $\chi_{imp}\sim -(T^2\ln T)^{-1}$)\cite{19,20}.
\par
Similar phenomena may exist in higher dimensions. The antiferromagnetic Kondo coupling indicates
a local potential well for the magnons. Therefore, some bound states of the magnons may exist in the
ground-state-configuration, which indicate the formation of the local spin-singlet. In this sense,
the impurity behaves as a diamagnetic center. When $J<0$, the Kondo coupling provides a repulsive
potential to the magnons and no local bound state can exist at low energy scales. The impurity must be
locked into the bulk. \par
In conclusion, we solve the model of a ferromagnetic Heisenberg chain coupled with a boundary impurity
with arbitrary spin. It is found that as long as the Kondo coupling is antiferromagnetic, (i) the impurity
spin behaves as a diamagnetic center and is completely screened by $2S$ bulk spins in the ground state, 
no matter how large the impurity spin 
is; (ii) The specific heat of the local composite (impurity plus $2S$ bulk spins which form bound state
with it) shows a simple power law $C_{loc}\sim T^{\frac32}$. We note that for a finite density of impurities, the local bound states are 
asymptotically extended to an impurity-band of the magnons, which is very similar to that of a ferrimagnetic
system. The critical behavior may be different from that of the single impurity case. When the impurity
density $n_i\sim 1/(2S)$, we expect a spin singlet ground state. 
\par
YW acknowledges the financial supports of AvH-Stiftung and NSF
of China. He is also indebted to the hospitality of  Institut f\"{u}r Physik, Universit\"{a}t Augsburg.


\begin{references}
\bibitem{22}Notice the leading term of the free energy upon the ground-state energy comes from 
the small $\lambda$ corrections to the strong coupling solution, which is mainly determined by
the weak coupling solution as shown in (9). In fact, the bulk strings serve as a variety of spin waves
 with a generalized exclusion statistics. This may explain why the spin-wave theory give the 
 correct critical exponents of the ferromagnetic Heisenberg chain [20,21].
\end{references}
\end{document}